\documentclass[letter]{aa} 
\usepackage{dcolumn}

\usepackage{graphicx}
\usepackage{txfonts}
\newcommand{\Hb}{H$\beta$}

\newcommand{\sVl}[3]{#1\,{\sc #2}]\,$\lambda{#3}$}
\newcommand{\Nl}[3]{#1\,{\sc #2}\,$\lambda{#3}$}
\makeatletter
 \def\hlinewd#1{%
   \noalign{\ifnum0=`}\fi\hrule \@height #1 \futurelet
    \reserved@a\@xhline}
\makeatother
\newcommand{\htopline}{\hlinewd{.8pt}}
\newcommand{\hmidline}{\hlinewd{.2pt}}
\newcommand{\hbotline}{\htopline}
\newcommand{\mcc}[1]{\multicolumn{1}{c}{#1}}

\newcommand{\mcr}[1]{\multicolumn{1}{r}{#1}}
\newcommand{\kms}{km\,s$^{-1}$}

\begin{document}

   \title{Accretion disk wind as explanation for the broad-line region
           structure in NGC\,5548}

   \author{W. Kollatschny 
          ,  M. Zetzl 
          }

   \institute{Institut f\"ur Astrophysik, Universit\"at G\"ottingen,
              Friedrich-Hund Platz 1, D-37077 G\"ottingen, Germany\\
              \email{wkollat@astro.physik.uni-goettingen.de}
}

   \date{Received December 14, 2012; accepted January 26, 2013}
   \authorrunning{Kollatschny, Zetzl}
   \titlerunning{Broad-line region structure in NGC\,5548}

 
  \abstract
   {Supermassive black holes in the centers of active galactic nuclei (AGN)
   are surrounded by broad-line regions (BLRs). The broad emission lines
   seen in the AGN spectra are emitted in this spatially unresolved region.}
   {We intend to obtain information on the structure and geometry
    of this BLR based on observed line profiles.}
   {We modeled the rotational and turbulent velocities in the
    line-emitting region on the basis of the
   line-width FWHM and line dispersion $\sigma_{line}$ of the variable
    broad emission lines in NGC\,5548.
    Based on these velocities we estimated the height
    of the line-emitting regions above the midplane
    in the context of their distances
    from the center.}
   {The broad emission lines originate at distances of 2 to 27 light days from
    the center. 
    Higher ionized lines originate in the inner region ($\le$ 13 light days)
    in specific filamentary structures
    1 to 14 light days above the midplane.
    In contrast, the H$\beta$ line is emitted in an outer
    (6 - 26 light days), more flattened
     configuration at heights of 0.7 to 4 light days only above the midplane.}
   {The derived geometry
   of the line-emitting region
    in NGC\,5548
     is consistent with an outflowing wind
    launched from an accretion disk.}


   \maketitle
%

\section{Introduction}

It is now generally accepted that active galactic nuclei (AGN)
are powered by accretion onto a super-massive black hole.
The broad emission lines we observe in the UV/optical regime
are generated by photoionization
in the outer regions of an accretion disk that surrounds
the central black hole.
Many details of this line-emitting region are unknown.
The broad-line region with an extension of about ten light days is
spatially unresolved on direct images.
However, some basic information about the distances of the line-emitting
regions from the central ionizing region
 can be obtained from reverberation mapping
(e.g. Clavel et al.\citealt{clavel91}, Peterson et al.\citealt{peterson04}),
i.e., the delayed variability of the integrated emission lines with respect
to that of the ionizing continuum.
Furthermore, there is stratification in the broad-line region. The higher
ionized lines originate closer to the central ionizing source than the
lower ionized lines. In a few cases the individual delays of emission
line segments (velocity delay maps) could be studied. Comparing these
velocity delay maps with model calculations point to the existence
of accretion disks with additional signatures of accretion disk
winds (Kollatschny\citealt{kollatschny03}, Bentz et al.\citealt{bentz10}).

Little is known about the size and geometry of the broad-line region
perpendicular to the accretion disk.
There are many models dealing with the geometry and structure of accretion
disks in AGN, as well as accretion disk winds
(e.g. Blandford\citealt{blandford82},
Collin-Souffrin et al.\citealt{collin88},
 Emmering et al.\citealt{emmering92},  
 K\"onigl \& Kartje\citealt{koenigl94}, DeKool \& Begelman\citealt{dekool95},
 Murray \& Chiang\citealt{murray97},
\citealt{murray98}, Bottorff et al.\citealt{bottorff97}, 
Blandford \& Begelman\citealt{blandford99},
Elvis\citealt{elvis00},
Proga, Stone \& Kallman\citealt{proga00}, Proga \& Kallman\citealt{proga04},
 Kollatschny\citealt{kollatschny03}, \citealt{kollatschny13},
Ho\citealt{ho08}, Goad et al.\citealt{goad12} and references therein).
The origin of an accretion disk winds is explained by radiation-driven
winds or magnetocentrifugal winds.

We have demonstrated 
in two recent papers (Kollatschny \& Zetzl\citealt{kollatschny11},
hereafter called Paper I, and Kollatschny \& Zetzl\citealt{kollatschny13},
hereafter called Paper II)
that general relations exist between 
the full-width at half maximum (FWHM) and
 the line-width ratio 
FWHM/$\sigma_{line}$ in the broad emission lines of AGN.
 The line-width FWHM reflects the
rotational motion of the broad-line gas in combination with an associated
turbulent motion. This turbulent velocity is different for the
different emission lines.
%
The rotational and turbulent velocities
give us information on the accretion disk
height with respect to the accretion disk radius of the line-emitting regions.
We know the absolute numbers of the line-emitting radii from reverberation
mapping, so we can get information on the absolute heights of
 the line-emitting regions
above the accretion disks.
 Here we present results for the broad-line region
geometry of NGC~5548.


\section{The NGC~5548 data sample}


One of the most extensive studied Seyfert galaxies is NGC~5548.
The spectra of the broad optical emission lines, including the
 H$\beta$ line, have been monitored over more than ten years
by large international
collaborations (e.g. Peterson et al.\citealt{peterson02} and references
therein).

Furthermore, two additional campaigns have been carried out in combination
with the IUE and HST satellites.
In a first combined optical/UV variability campaign NGC~5548 was
monitored with the IUE satellite for a period of eight months from 1988 December
until 1989 August (Clavel et al.\citealt{clavel91}), as well as in the optical
from 1988 December until 1989 October (Peterson et al.\citealt{peterson91}).
In a second combined optical/UV variability campaign NGC~5548 was
monitored with the IUE and HST, along with ground-based telescopes, for a period
from 1992 October until 1993 September (Korista et al.\citealt{korista95}).
 
Our current investigation is based on all spectral information of
the root-mean-square (rms) emission line profiles in NGC~5548
(Peterson et al.\citealt{peterson04}). The narrow line components
disappear in these spectra. This sample has been the basis for our
Papers I and II as well.

Altogether, we have information about the rms line profile widths FWHM
and $\sigma_{line}$, along with
distances of the emitting regions from the central ionizing source for the
following emission lines: the
optical H$\beta$ and \ion{He}{ii}\,$\lambda 4686$ lines, and the UV
\ion{He}{ii}\,$\lambda 1640$,
\ion{C}{iii}]\,$\lambda 1909$, \ion{C}{iv}\,$\lambda 1550$, and 
\ion{S}{iv}\,$\lambda 1400$ lines  (see Table 1).
For the  H$\beta$  line we know their annual rms profiles over
a period of 14 years (1988 - 2001) and their related distances.
 For the  \ion{He}{ii}\,$\lambda 4686$ 
line we only know one rms profile based on the campaign in 1988/1989.
However, for the rest of the UV lines we have two rms profiles
based on the campaigns in the years 1988/89 and 1992/93.


\section{Results}


\subsection{Observed and modeled emission line-width
 ratios}

We parameterize the rms
line profiles by both their FWHM and the
ratio of
their FWHM to their line dispersion
$\sigma_{line}$. 
We present in Table 1 and Figs.\,1 to 3 the observed line widths of the 
emission lines in NGC\,5548, the corresponding modeled
turbulent $v_{turb}$, and 
rotational velocities $v_{rot}$ of the line-emitting regions (see Papers I, II).
The $\sigma_{line}$ values of the modeled profiles are integrated
over line-widths ~25,000 km\,s$^{-1}$ (see Paper I). The modeled line-width
ratios  FWHM/$\sigma_{line}$, hence the turbulent velocities, would
decrease/increase by 10 - 20 percent if we integrated over line-widths that
are broader/smaller by 20 percent. However, the general trends remain the same.

\begin{table*}[htbp]
    \centering
       \leavevmode
       \tabcolsep1mm
        \newcolumntype{d}{D{.}{.}{-2}}
        \newcolumntype{p}{D{+}{\,\pm\,}{-1}}
        \newcolumntype{K}{D{,}{}{-2}}

\caption{Line profile parameters and the line-emitting regions
of the individual emission lines.
}
\begin{tabular}{lKKKKKKdKd}
 \htopline
\hspace{3mm} Line &\mcr{FWHM }& \mcr{FWHM/$\sigma$} &\mcc{v$_{\text{turb}}$}&\mcc{v$_{\text{rot}}$} &\mcc{Radius}
                  & \mcc{Height} & \mcr{H/R} & \mcc{Height$_{\text{corr}}$} & \mcc{H$_{\text{corr}}$/R}\\
\hspace{3mm}      &\mcr{[\kms{}]}&&\mcc{[\kms{}]}&\mcc{[\kms{}]} &\mcc{[ld]}
                  & \mcc{[ld]} & \mcr{} & \mcc{[ld]} & \mcr{}\\
 \hmidline
\multicolumn{3}{l}{\hspace{-3mm}NGC 5548 -- 1988/1989 (opt+IUE)}\\
 \hmidline
\hspace{3mm}\Nl{He}{ii}{1640}  &9803,\pm{1594}&2.516,\pm{0.130}&1943,^{+1763}_{-1383} &5700,^{+974}_{-1356}&3.8,^{+1.7}_{-1.8}&1.3,\pm{1.4}&0.34&1.5,\pm{1.4}&0.39\\
\hspace{3mm}\Nl{He}{ii}{4686}  &7338,\pm{901}&2.575,\pm{0.103}&928,^{+573}_{-536} &4297,^{+528}_{-579}&7.8,^{+3.2}_{-3.0}&1.7,\pm{1.3}&0.22&1.6,\pm{1.3}&0.21\\
\hspace{3mm}\Nl{C}{iv}{1549}   &6556,\pm{878}&1.706,\pm{0.076}&3537,^{+1256}_{-1050} &3263,^{+901}_{-1492}&9.8,^{+1.9}_{-1.5}&10.6,\pm{6.5}&1.08&8.7,\pm{5.7}&0.89\\
\hspace{3mm}\Nl{Si}{iv}{1400}  &6455,\pm{3030}&2.506,\pm{0.288}&807,^{+1489}_{-10806} &3780,^{+1737}_{-2287}&12.3,^{+3.4}_{-3.0}&2.6,\pm{35.2}&0.21&6.8,\pm{35.5}&0.55\\
\hspace{3mm}\Hb                &4044,\pm{199}&2.397,\pm{0.061}&374,^{+86}_{-85} &2368,^{+116}_{-117}&19.7,^{+1.5}_{-1.5}&3.1,\pm{0.8}&0.16&3.3,\pm{0.8}&0.17\\
\hspace{3mm}\sVl{C}{iii}{1909} &5018,\pm{1458}&2.126,\pm{0.156}&907,^{+657}_{-646} &2925,^{+867}_{-1028}&27.4,^{+5.4}_{-5.3}&8.5,\pm{7.0}&0.31&14.1,\pm{8.4}&0.51\\
 \hmidline
\multicolumn{3}{l}{\hspace{-3mm}NGC 5548 -- 1992/1993 (opt+IUE+HST)}\\
 \hmidline
\hspace{3mm}\Nl{He}{ii}{1640}  &8929,\pm{1571}&2.031,\pm{0.056}&4137,^{+2222}_{-1662} &4691,^{+1370}_{-2539}&1.9,^{+0.3}_{-0.3}&1.7,\pm{1.3}&0.89&0.9,\pm{1.0}&0.47\\
\hspace{3mm}\Nl{Si}{iv}{1400}  &7044,\pm{1849}&1.755,\pm{0.090}&3878,^{+6121}_{-1789} &3474,^{+1661}_{-2429}&4.3,^{+1.1}_{-1.0}&4.8,\pm{8.4}&1.12&2.6,\pm{7.8}&0.60\\
\hspace{3mm}\Nl{C}{iv}{1549}   &6868,\pm{450}&2.064,\pm{0.051}&2002,^{+426}_{-400} &3906,^{+333}_{-377}&6.7,^{+0.9}_{-1.0}&3.4,\pm{1.0}&0.51&5.0,\pm{1.1}&0.75\\
\hspace{3mm}\Hb                &7202,\pm{392}&2.488,\pm{0.075}&1048,^{+336}_{-316} &4213,^{+235}_{-257}&13.4,^{+3.8}_{-4.3}&3.3,\pm{1.5}&0.25&1.3,\pm{1.1}&0.10\\
\hspace{3mm}\sVl{C}{iii}{1909} &4895,\pm{1263}&1.517,\pm{0.070}&2488,^{+811}_{-796} &2495,^{+1024}_{-1615}&13.9,^{+1.8}_{-1.4}&13.9,\pm{10.2}&1.00&8.4,\pm{7.1}&0.60\\
 \hmidline
\multicolumn{3}{l}{\hspace{-3mm}NGC 5548 -- H$\beta$}\\
 \hmidline
\hspace{3mm}\Hb &5957,\pm{224}&3.025,\pm{0.055}&223,^{+104}_{-102} &3468,^{+129}_{-128}&6.5,^{+5.7}_{-3.7}&0.4,\pm{0.4}&0.06&0.7,\pm{0.7}&0.11\\
\hspace{3mm}\Hb &8047,\pm{1268}&2.614,\pm{0.127}&1043,^{+879}_{-792} &4711,^{+743}_{-847}&7.8,^{+3.8}_{-2.8}&1.7,\pm{1.7}&0.22&0.7,\pm{1.5}&0.09\\
\hspace{3mm}\Hb &5691,\pm{164}&2.514,\pm{0.074}&614,^{+153}_{-146} &3333,^{+96}_{-99}&11.0,^{+1.9}_{-2.0}&2.0,\pm{0.6}&0.18&1.3,\pm{0.6}&0.12\\
\hspace{3mm}\Hb &7202,\pm{392}&2.488,\pm{0.075}&1048,^{+336}_{-316} &4213,^{+235}_{-257}&13.4,^{+3.8}_{-4.3}&3.3,\pm{1.5}&0.25&1.3,\pm{1.1}&0.10\\
\hspace{3mm}\Hb &6247,\pm{343}&2.875,\pm{0.088}&361,^{+194}_{-187} &3648,^{+199}_{-200}&14.3,^{+5.9}_{-7.3}&1.4,\pm{1.1}&0.10&1.6,\pm{1.1}&0.11\\
\hspace{3mm}\Hb &5776,\pm{237}&2.784,\pm{0.082}&377,^{+149}_{-143} &3376,^{+139}_{-140}&15.9,^{+2.9}_{-2.5}&1.8,\pm{0.8}&0.11&1.9,\pm{0.8}&0.12\\
\hspace{3mm}\Hb &5706,\pm{357}&2.816,\pm{0.071}&343,^{+160}_{-157} &3333,^{+206}_{-206}&16.4,^{+1.2}_{-1.1}&1.7,\pm{0.8}&0.10&2.0,\pm{0.8}&0.12\\
\hspace{3mm}\Hb &5541,\pm{354}&2.881,\pm{0.070}&280,^{+144}_{-143} &3233,^{+202}_{-204}&17.5,^{+2.0}_{-1.6}&1.5,\pm{0.8}&0.09&2.2,\pm{0.8}&0.13\\
\hspace{3mm}\Hb &4664,\pm{324}&2.478,\pm{0.083}&435,^{+152}_{-147} &2731,^{+189}_{-191}&18.6,^{+2.1}_{-2.3}&3.0,\pm{1.1}&0.16&2.7,\pm{1.1}&0.15\\
\hspace{3mm}\Hb &4044,\pm{199}&2.397,\pm{0.061}&374,^{+86}_{-85} &2368,^{+116}_{-117}&19.7,^{+1.5}_{-1.5}&3.1,\pm{0.8}&0.16&3.3,\pm{0.8}&0.17\\
\hspace{3mm}\Hb &6142,\pm{289}&2.733,\pm{0.123}&473,^{+244}_{-226} &3593,^{+171}_{-172}&21.7,^{+2.6}_{-2.6}&2.9,\pm{1.5}&0.13&2.4,\pm{1.5}&0.11\\
\hspace{3mm}\Hb &6377,\pm{147}&3.221,\pm{0.036}&127,^{+67}_{-67} &3700,^{+84}_{-84}&24.8,^{+3.2}_{-3.0}&0.9,\pm{0.5}&0.04&2.7,\pm{0.6}&0.11\\
\hspace{3mm}\Hb &4596,\pm{505}&2.654,\pm{0.088}&309,^{+168}_{-169} &2686,^{+291}_{-292}&26.5,^{+4.3}_{-2.2}&3.0,\pm{1.8}&0.11&3.9,\pm{1.8}&0.15\\
\hbotline
\end{tabular}
\end{table*}
The observed and modeled line-width ratio
 FWHM/$\sigma_{line}$ versus line-width FWHM are presented separately for the
two variability campaigns of the years 1988/89 and 1992/93, as well as
for the 14 H$\beta$ rms lines based on their annual profiles
for the years 1988 until 2001.
%
%
\begin{figure}
\includegraphics[width=6.3cm,angle=270]{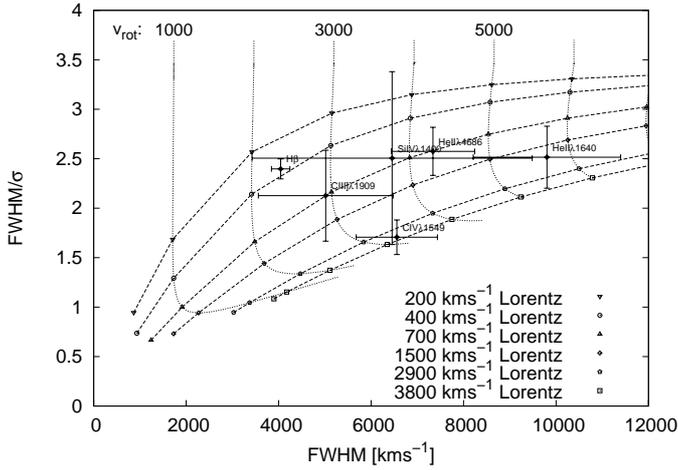} 
       \vspace*{5mm} 
  \caption{NGC~5548: Observed and modeled line-width ratios
 FWHM/ $\sigma_{line}$ versus line-width FWHM for the period 1988/89.}
   \label{ngc5548_time1.ps}
\end{figure}
%
%
%
\begin{figure}
\includegraphics[width=6.3cm,angle=270]{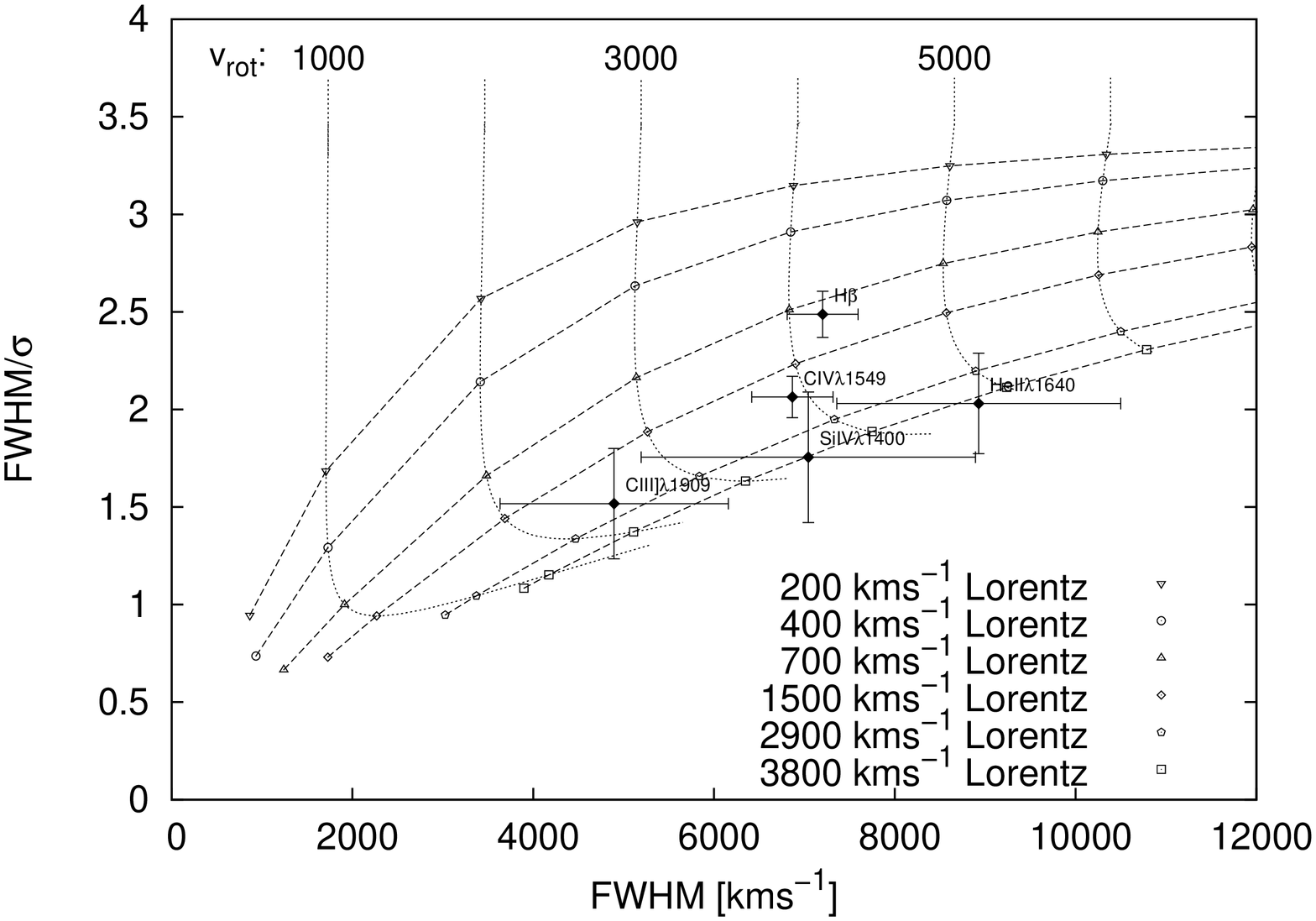} 
       \vspace*{5mm} 
  \caption{NGC~5548:
 Observed and modeled line-width ratios
 FWHM/ $\sigma_{line}$ versus line-width FWHM for the period 1992/93.}
   \label{ngc5548_time2.ps}
\end{figure}
%
%
%
\begin{figure}
\includegraphics[width=6.3cm,angle=270]{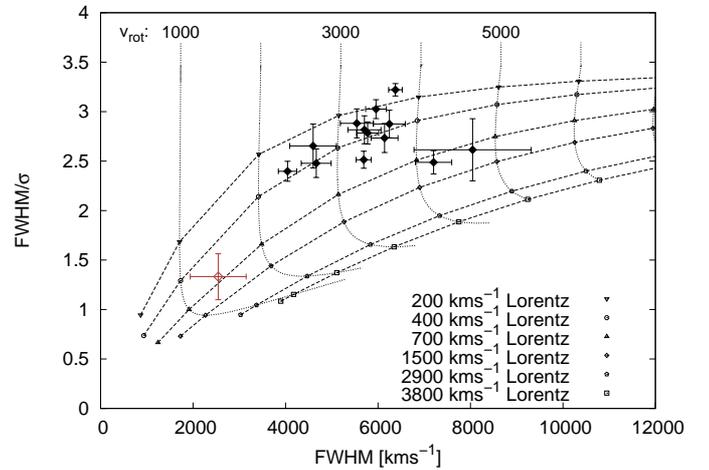} 
       \vspace*{5mm} 
  \caption{NGC~5548: Observed and modeled $H\beta$ line-width ratios
 FWHM/ $\sigma_{line}$ versus line-width FWHM over 14 years (1988 - 2001).}
   \label{ngc5548_hbeta.ps}
\end{figure}
%
%
One H$\beta$ rms profile strongly deviates from the H$\beta$ profiles
 of the other years
in Fig. 3 (red cross at FWHM = 2500\,kms$^{-1}$). This profile  
has been considered to be less reliable by 
Peterson et al.\cite{peterson04} before. We neglect this individual profile
for the rest of our investigation.

The ratio of the turbulent velocity $v_{turb}$ over the
rotational velocity $v_{rot}$ in the line-emitting region
gives us information on the ratio of the accretion disk
height  $H$ with respect to the accretion disk radius $R$
 of the line-emitting regions
as presented in Papers I and II:
\begin{equation}
\label{eq:HtoR}
  H/R = (1/\alpha) (v_{turb}/v_{rot}). 
\end{equation}
The unknown viscosity parameter $\alpha$ is assumed to be constant
and to have a value of one. In reality, the value of $\alpha$ might be up to
one order lower.

Since we know the distances $R$ 
of the line-emitting regions from reverberation
mapping (Table~1), we are able to estimate the height $H$ of the
line-emitting region. We present in Table~1 information on both the height of
the line-emitting region in units of light days and the
ratio $H/R$. 

\subsection{Broad-line region geometry of NGC\,5548}

The broad-line region structure of NGC~5548
based on the radius and height data in Table\,1 is shown in Fig.~4.
Given are the emitting regions of the  \ion{He}{ii}\, $\lambda 1640$, 
\ion{Si}{iv}\,$\lambda 1400$,
\ion{C}{iv}\,$\lambda 1550$, \ion{C}{iii}]\,$\lambda 1909$, as well as
 H$\beta$  rms emission lines (red symbols) as a function of distance
to the center and of the height above the midplane
for the two epochs 1988/89 (1) and 1992/93 (2).
Furthermore, the position of the H$\beta$  emitting region
is presented for all 13 rms spectra obtained for
the years 1988 to 2001 (H$\beta$ all).
The \ion{He}{ii}\,$\lambda 4686$ line has only been monitored at one epoch.
The dot at radius zero gives the size of a Schwarzschild black hole (with
$M=6.7\times10^{7}M_{\sun}$) multiplied by a factor of twenty.
The two axes' scale in Fig.~4 are linear in units of light days.
One light day corresponds to a distance of 2.59 $\times10^{15}$cm.
The axis on top of the figure gives the distance of the line-emitting regions
from the center in units of the
Schwarzschild radius (for a central black hole mass
of $ M=6.7\times10^{7}M_{\sun}$ taken from 
Peterson et al.\citealt{peterson04}).
We make the assumption that the accretion disk structure is arranged
symmetrically to the midplane in Figs. 4 and 5.
%
\begin{figure}
\includegraphics[width=8.0cm,angle=0]{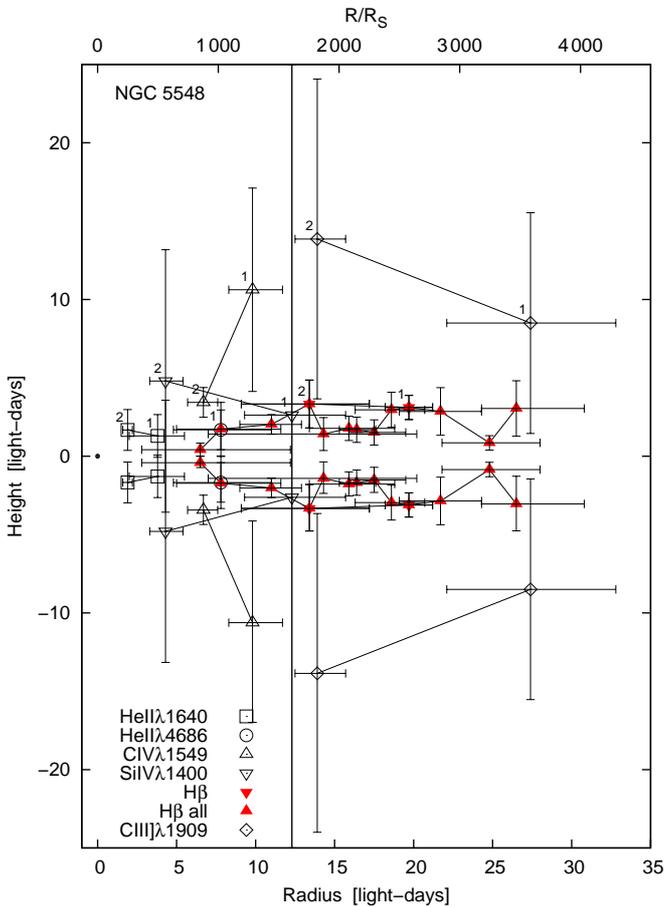} 
  \caption{Broad-line region structures of NGC~5548
based on the dominant emitting regions of the 
broad optical/UV lines as a function of distance
to the center and of height above the midplane.
The emission regions of the individual lines that are observed
at different epochs are connected by a solid line.
The dot at radius zero gives the size of a Schwarzschild black hole (with
$M=6.7\times10^{7}M_{\sun}$) multiplied by a factor of twenty.}
   \label{disc_ngc5548_t1t2hb_layer_old.eps}
\end{figure}
%

The errors of the line widths (in Table 1) and therefore of the 
derived turbulent velocities  $v_{turb}$ 
are quite large for the weak UV emission lines in NGC~5548.
This especially applies
to the \ion{Si}{iv}\,$\lambda 1400$ and
 \ion{C}{iii} ]\,$\lambda 1909$ line profiles based on IUE spectra.
We demonstrated in Papers I and II that dedicated turbulent velocities 
belong to the individual emission line regions.
These dedicated velocities have been derived
from many line profiles.   
Therefore, we then calculated additional corrected heights of the line-emitting
regions based on the turbulent velocities belonging to  the individual lines
(see Papers I, II) : 400 km\,s$^{-1}$ for H$\beta$,  
900 km\,s$^{-1}$ for \ion{He}{ii}\,$\lambda 4686$,
1500 km\,s$^{-1}$ for \ion{C}{iii}]\,$\lambda 1909$,
2100 km\,s$^{-1}$ for  \ion{Si}{iv}\,$\lambda 1400$,
2300 km\,s$^{-1}$ for \ion{He}{ii}\,$\lambda 1640$, and
2900 km\,s$^{-1}$ for \ion{C}{iv}\,$\lambda 1549$.
We give in Table 1 the corrected
height Height$_{\text{corr}}$ of the line-emitting regions based on
these $v_{turb}$.
The broad-line region structure of
NGC~5548 based on the corrected turbulent
velocities $v_{turb}$ is shown in Fig.~5.

%
\begin{figure}
\includegraphics[width=8.0cm,angle=0]{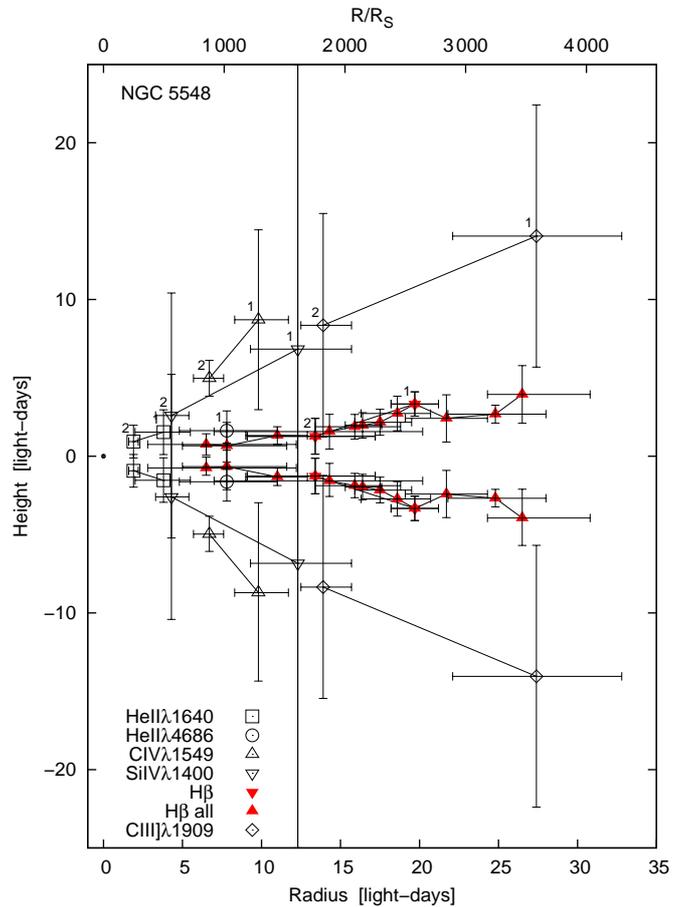} 
  \caption{Broad-line region structure of
NGC~5548 based on the dominant emitting regions of the 
broad optical/UV lines as a function of distance
to the center as well as height above the midplane.
This figure is based on corrected turbulent
velocities $v_{turb}$.  The dot at radius zero gives the size of a
Schwarzschild black hole (with $M=6.7\times10^{7}M_{\sun}$) multiplied by a
factor of twenty.}
   \label{disc_ngc5548_t1t2hb_layer.eps}
\end{figure}
%

The radius of the dominating H$\beta$ line-emitting region varied by 
a factor of more than four during the monitoring campaign from the year
 1988 until 2001.
It has been shown before that the distances of the line-emitting regions
depend on the luminosity of the central ionizing source
(e.g. Dietrich \& Kollatschny\citealt{dietrich95},
Peterson et al.\citealt{peterson02},\citealt{peterson04}).
It should be emphasized that the individual emission lines do not originate
at one single radius only, but rather in an extended region 
(see e.g.  Kollatschny\citealt{kollatschny03}). Therefore it is reasonable
to connect the individual line-emitting regions as shown in
Figs.~4 and 5.
Furthermore,
it was known before that higher ionized lines originate closer to
the central ionizing source as seen in Fig.5. The 
\ion{C}{iv}\,$\lambda 1549$ line, e.g., originates inwards of the
\ion{C}{iii}]\,$\lambda 1909$ line.

The H$\beta$ lines are emitted in a more flattened configuration
above the midplane.  
The H$\beta$ line originates at a height
of 0.7 light days only at a radius of seven light days.
This corresponds to theoretical $H/R$ values of 0.01 -- 0.3
(e.g. DeKool \& Begelman\citealt{dekool95})
 based on accretion disk models .
However, the higher ionized lines in NGC~5548 originate in a far more
extended region
above the presumed accretion disk. We observe H/R values of 0.1 until
0.9 (Table 1, Fig.5) for the line-emitting regions.
This indicates that the emission lines do not originate in a thin
atmosphere of an accretion disk but rather in filaments at greater
heights above the disk.
The different geometries of the high/low ionization lines might be explained by
a nonspherical geometry of the photoionizing source.

The observed geometry of the BLR in NGC~5548 strikingly corresponds to the
disk wind models of
Murray \& Chiang (\!\!\citealt{murray97}, their Fig. 1)
and Proga \& Kallman (\!\!\citealt{proga04}, their Fig. 1d).
Furthermore, the emitting region of the H$\beta$ line is arranged more
horizontally 
in comparison to the higher ionized lines.
It has been predicted by Murray \& Chiang \cite{murray97} in their models 
that the angle the streamlines make with the disk vary with the
distance/radius of the footprint of the streamline. The
streamlines --- based on their model --- should be more vertical at 
smaller radii, as seen in Fig.~5. 


\section{Conclusions}

    We demonstrate in our investigation that   
    the higher ionized lines of the broad-line region originate in an
    extended region of 1 to 14 light days above the midplane.
    In contrast, the H$\beta$ line only originates at distances of
    0.7 to 4 light days above the midplane.
   The derived filamentary geometry of the broad-line emitting region
   in NGC\,5548 is consistent with models of an outflowing wind
   launched from an accretion disk.

\begin{acknowledgements}
      Part of this work was supported by the German
      \emph{Deut\-sche For\-schungs\-ge\-mein\-schaft, DFG\/} project
      number Ko~857/32-1.
\end{acknowledgements}


\begin{thebibliography}{}

 \bibitem[\protect\citeauthoryear{}{2010}]{bentz10} Bentz, M.,~C. et al. 2010, ApJ, 720, L46
%
 \bibitem[\protect\citeauthoryear{}{1982}] {blandford82} Blandford, R.D. \& Payne, D.G. 1982, MNRAS,
  199, 883
%
 \bibitem[\protect\citeauthoryear{}{1999}] {blandford99} Blandford, R.D. \& Begelman, M.C. 1999, MNRAS,
  303, L1
%
 \bibitem[\protect\citeauthoryear{}{1997}]{bottorff97} Bottorff, M.,~C. et al. 1997, ApJ, 479, 200
%
%
%
 \bibitem[\protect\citeauthoryear{}{1991}]{clavel91} Clavel, J. et al. 1991, ApJ, 366, 64
%
 \bibitem[\protect\citeauthoryear{}{1988}]{collin88} Collin-Souffrin, S., Dyson, J.~E., McDowell, J.~C.
\& Perry, J.~J. 1988, MNRAS, 232, 539
%
  \bibitem[\protect\citeauthoryear{}{2006}]{collin06} Collin, S., Kawaguchi, T., Peterson, B.~M.,
 Vestergaard, M. 2006, A\&A, 456, 75
%
  \bibitem[\protect\citeauthoryear{}{1995}] {dekool95} DeKool, M. \& Begelman M.C. 1995, ApJ, 455, 448 
%
   \bibitem[\protect\citeauthoryear{}{1995}] {dietrich95} Dietrich, M. \& Kollatschny, W. 1995, A\&A,
 303, 405 
%
  \bibitem[\protect\citeauthoryear{}{2000}]{elvis00} Elvis, M. 2000, ApJ, 545, 63
%
  \bibitem[\protect\citeauthoryear{}{1992}]{emmering92} Emmering, R.T., Blandford, R.D., \& Shlosman, I.
       1992, ApJ, 385, 460
%
  \bibitem[\protect\citeauthoryear{}{2012}]{goad12} Goad, M.~R., Korista, K.~T. \& Ruff, A.~J. 2012,
    MNRAS, 426, 3086
%
 \bibitem[\protect\citeauthoryear{}{2008}]{ho08} Ho, L. 2008, ARA\&A, 45, 475 
%
  \bibitem[\protect\citeauthoryear{}{1994}]{koenigl94} K\"{o}nigl, A., \& Kartje, J.E.
       1994, ApJ, 434, 446
%
  \bibitem[\protect\citeauthoryear{}{2003}]{kollatschny03} Kollatschny, W. 2003, A\&A, 407, 461
%
%
  \bibitem[\protect\citeauthoryear{}{2011}]{kollatschny11} Kollatschny, W. \& Zetzl, M. 2011,
  Nature, 470, 366  (Paper I)
%
 \bibitem[\protect\citeauthoryear{}{2013}]{kollatschny13} Kollatschny, W. \& Zetzl, M. 2013,
  A\&A, 549, A100 (Paper II)
%
 \bibitem[\protect\citeauthoryear{}{1995}]{korista95} Korista, K.~T. et al. 1995, ApJS, 97, 285
%
 \bibitem[\protect\citeauthoryear{}{1997}]{murray97} Murray, N., \& Chiang, J. 1997, ApJ, 474, 91
%
\bibitem[\protect\citeauthoryear{}{1998}]{murray98} Murray, N., \& Chiang, J. 1998, ApJ, 494, 125
%
 \bibitem[\protect\citeauthoryear{}{1991}]{peterson91} Peterson, B.~M. et al. 1991,
ApJ, 368, 119 
%
 \bibitem[\protect\citeauthoryear{}{2002}]{peterson02} Peterson, B.~M. et al. 2002,
ApJ, 581, 197 
%
 \bibitem[\protect\citeauthoryear{}{2004}]{peterson04} Peterson, B.~M. et al. 2004,
ApJ, 613, 682
%
%
\bibitem[\protect\citeauthoryear{}{2000}]{proga00} Proga, D., Stone, J.M. \& Kallman, T.R. 2000,
ApJ, 543, 686 
%
\bibitem[\protect\citeauthoryear{}{2004}]{proga04} Proga, D. \& Kallman, T.R. 2004,
ApJ, 616, 688 
%
%
\end{thebibliography}
\end{document}